\documentclass[11pt]{article}
\usepackage{tabularx,booktabs}
\usepackage{url}
\usepackage{fullpage}
\usepackage{graphicx}
\usepackage{lipsum}
\usepackage{amsmath}
\usepackage{amsthm}
 \usepackage{color}
 \usepackage{geometry}
\begin{document}
\thispagestyle{empty}

\title{scDiffusion: conditional generation of high-quality single-cell data using diffusion model}

\author{  
Erpai Luo$^{1,^\#}$, Minsheng Hao$^{1,^\#}$, Lei Wei$^{1}$, Xuegong Zhang$^{1,2,^*}$\\
\\ 
\fontsize{10pt}{\baselineskip}\selectfont $^{1}$MOE Key Lab of Bioinformatics and Bioinformatics Division of BNRIST, \\ 
\fontsize{10pt}{\baselineskip}\selectfont Department of Automation, Tsinghua University, Beijing 100084, China\\ 
\fontsize{10pt}{\baselineskip}\selectfont $^{2}$School of Life Sciences and School of Medicine, Tsinghua University, Beijing 100084, China\\	 
}

\date{}

\renewcommand{\thefootnote}{\fnsymbol{footnote}}  
\footnotetext{$^\#$ These authors contributed equally to this work.}  
\footnotetext{$^*$ Corresponding Author. Email: zhangxg@tsinghua.edu.cn} 

\maketitle

\begin{abstract}

Single-cell RNA sequencing (scRNA-seq) data are important for studying the laws of life at single-cell level. However, it is still challenging to obtain enough high-quality scRNA-seq data. To mitigate the limited availability of data, generative models have been proposed to computationally generate synthetic scRNA-seq data. Nevertheless, the data generated with current models are not very realistic yet, especially when we need to generate data with controlled conditions. In the meantime, the Diffusion models have shown their power in generating data at high fidelity, providing a new opportunity for scRNA-seq generation.

In this study, we developed scDiffusion, a generative model combining diffusion model and foundation model to generate high-quality scRNA-seq data with controlled conditions. We designed multiple classifiers to guide the diffusion process simultaneously, enabling scDiffusion to generate data under multiple condition combinations. We also proposed a new control strategy called Gradient Interpolation. This strategy allows the model to generate continuous trajectories of cell development from a given cell state.

Experiments showed that scDiffusion can generate single-cell gene expression data closely resembling real scRNA-seq data. Also, scDiffusion can conditionally produce data on specific cell types including rare cell types. Furthermore, we could use the multiple-condition generation of scDiffusion to generate cell type that was out of the training data. Leveraging the Gradient Interpolation strategy, we generated a continuous developmental trajectory of mouse embryonic cells. These experiments demonstrate that scDiffusion is a powerful tool for augmenting the real scRNA-seq data and can provide insights into cell fate research. 
\end{abstract}

\newpage

\section{Introduction}

Single-cell RNA sequencing (scRNA-seq) data offer comprehensive depictions of the gene expression profile of every single cell,
gaining a more systematic and precise understanding of the development and function of living organisms \cite{jovic2022single, gohil2021applying}.
Although current sequencing technologies have come a long way, the cost and difficulty of sequencing remain high. Besides, the biological samples are sometimes hard to obtain \cite{jiang2022big,ke2022single,suva2019single}, and certain cell types within a sample may be too rare to be analyzed. 
It is still challenging to obtain enough high-quality scRNA-seq data of interest, which may impede biological discovery as most tools for scRNA-seq analysis require a certain amount of high-quality data.

Some researchers have endeavored to generate in silico gene expression data that obviate the need for further biological samples, thus mitigating the limited availability of scRNA-seq data. This pseudo data is designed to meet specific criteria, thereby facilitating more effective downstream analysis. There are two main types of in silico data generation methods: statistical models and deep generative models. 
Statistical models use some well-studied statistical distributions such as Zero-inflated Negative Binomial (ZINB) \cite{greene1994accounting} to model the expression data, and new data are generated by manually setting certain parameters of the distributions \cite{lindenbaum2018geometry, dibaeinia2020sergio, li2019statistical, zappia2017splatter}. These manual designs may oversimplify the complex patterns in real scenarios, making these methods hardly mimic the real gene expression data well. Nowadays these statistical models are mainly used to generate toy data for guiding the development of scRNA-seq analysis algorithms.

The recent prosperity of deep generative models brings great chances for the in silico transcriptomic data generation \cite{lopez2020enhancing}. By principle, the current models can be summarized into two types: the variational autoencoder (VAE) based and the generative adversary network (GAN) based. Although the VAE-based models such as scVI \cite{lopez2018deep} are the most prominent in this field \cite{kingma2013auto}, these models are mainly focused on downstream analysis tasks (e.g. batch correction and clustering), rather than generating gene expression profiles of cells.
GAN-based models such as scGAN were proposed for generating new cells and accomplishing downstream tasks \cite{marouf2020realistic,lall2022lsh,xu2020scigans}. 
However, these GAN-based models could only generate data from a known distribution, unsatisfying the need to supplement unmeasured data. Besides, GAN requires careful designing and tuning of model architectures, as well as optimization tricks, to achieve a stable training process \cite{saxena2021generative,yang2022diffusion}, hindering the smooth application on new datasets and the generation of data under certain conditions.

Recently, the latent diffusion model (LDM) \cite{rombach2022high} has demonstrated excellent performance in several areas such as images, audio, and videos \cite{yang2022diffusion,cao2022survey,croitoru2023diffusion}. Compared with the GAN model, it has a stable training process and can easily generate samples conditioned on complex prompts\cite{dhariwal2021diffusion,bond2021deep,zhang2023text}. However, few studies have deployed it in the single-cell area. One of the challenges lies in the fact that LDM needs a pre-trained autoencoder model to link the data in the latent and original space. There are no such models in the single-cell field until the recent emergence of the foundation models \cite{theodoris2023transfer, bian2024scmulan, cui2023scgpt, hao2023large}. By using massive parameters and being trained on extensive datasets, these models can learn the unified representation of the gene expression data, which facilitates a variety of downstream tasks and can be used as the autoencoder model in LDM. 

We propose scDiffusion, an in silico scRNA-seq data generation model combining LDM with the foundation model, to generate single-cell gene expression data with given conditions. 
scDiffusion has three parts, an autoencoder, a denoising network, and a condition controller. 
We used the pre-trained model SCimilarity \cite{heimberg2023scalable} as an autoencoder to rectify the raw distribution and reduce the dimensionality of scRNA-seq data, which can make the data amenable to diffusion modeling. 
The denoising network was redesigned based on a skip-connected multilayer perceptron (MLP) to learn the reversed diffusion process.
The conditional controller is a cell type classifier, enabling scDiffusion to generate data specific to a particular cell or organ type according to diverse requirements.
We conducted a series of experiments to evaluate the performance of our model. First, we evaluated the single-conditional generation ability of scDiffusion. We generated new data on three different datasets and evaluated the generated data with different metrics. The results showed that the scDiffusion could generate realistic scRNA-seq data and had superior conditional generation ability.
Then we assessed the multi-conditional generation capabilities of scDiffusion. We deployed two separate classifiers to guide the generation. The results showcased the model's ability to generate out-of-distribution data through the amalgamation of known conditions.
We also proposed a new condition control strategy, Gradient Interpolation, to interpolate continuous cell trajectories from discrete cell states. We used this strategy to reconstruct the intermediate states within an embryonic development process. The results showed that the scDiffusion could bridge the gaps between sequencing intervals and provide a more comprehensive developmental timeline.
With the powerful generation ability, scDiffusion has the potential to augment existing scRNA-seq data and could potentially contribute to the investigation of undersampled or even unseen cell states.



\section{Methods}
\begin{figure}[h]
  \begin{center}
    \includegraphics[width=0.95\textwidth]{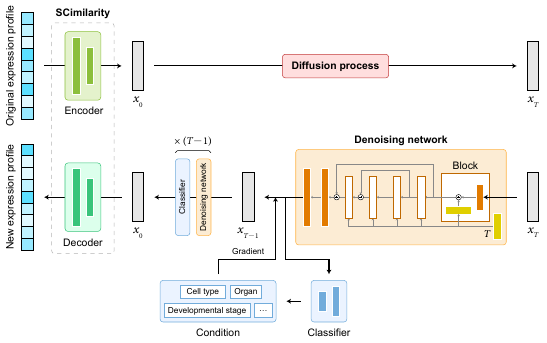}
  \end{center}
\caption{The overall structure of scDiffusion.}
\label{fig:model_structure}
\end{figure}

The scDiffusion model consists of three parts, a pre-trained foundation model SCimilarity \cite{heimberg2023scalable} as the autoencoder, a denoising network, and a conditional classifier, as depicted in Fig. \ref{fig:model_structure}.
At the training stage, the SCimilarity model is first fine-tuned based on the pre-trained weight to embed the gene expression profile. 
After that, the diffusion process is applied to each embedding derived by the autoencoder and produces a series of noisy embeddings. 
These noisy embeddings serve as the training data for the backbone network. 
Meanwhile, the conditional classifier processes the embeddings to predict associated labels, such as cell types. 
At the inference stage, the denoising network denoises the input noise embeddings and generates new embeddings. 
The generation can be guided by the classifier or the Gradient Interpolation strategy. 
The generated embeddings are finally fed into the decoder to obtain full gene expression. Detailed descriptions of scDiffusion are provided below.

\subsection{Finetuning the pre-trained foundation model as autoencoder}

An important prerequisite for LDM is to have a suitable autoencoder to concatenate the data in latent space and original space. To satisfy this prerequisite, we used the pre-trained foundation model SCimilarity \cite{heimberg2023scalable} as the autoencoder, encoding the gene expression data of every single cell $S_{ori}$ into a latent space embedding $x_0$. SCimilarity is an encoder-decoder style network trained on a 22.7 million cell corpus assembled across 399 published scRNA-seq studies and could be used to extract unifying representation from cell expression profiles.

We finetuned the pre-trained model weights with our data and used SCimilarity's encoder and decoder separately. The input of the encoder is a gene expression profile that is normalized by 1e4 total counts and then logarithmized, and the output is a 128-dimension latent space embedding as it was set in the pre-trained model. 
The decoder subsequently accepts the latent space embedding and generates the corresponding expression profile $S_{new}$ as the output. Leveraging the powerful representation and generalization capabilities of foundation model, we could extract complete information from cellular expression into latent representations and accurately rebuild the original gene expression.

During the fine-tuning process, we used the pre-trained weight except for the first layer and the last layer since the numbers of genes in our data differ from that in the pre-trained dataset. Compared to training a new autoencoder from scratch, this method gives faster and better access to the desired autoencoder (Fig \ref{fig:result_noncondi}a). As shown in Fig. \ref{fig:dis_trans}, the distribution of gene expression is transformed into a Gaussian-like distribution by the autoencoder, which is in line with the Gaussian distribution used in the diffusion process and makes it much easier for the denoising network to learn the reverse process. 

\subsection{Training the denoising network}

After getting embeddings from the encoder, the diffusion process is applied to each embedding to obtain noisy data. 
The denoising network is trained on these noisy embeddings to learn the reversed process. 
Classical denoising network models such as convolutional neural networks are not applicable to gene expression, as a gene expression profile of scRNA-seq data is a long, sparse, and unordered vector.
Thus, we developed a new architecture as the backbone, with fully connected layers and a skip-connected structure (Fig. \ref{fig:model_structure}). The skip-connected structure can help to maintain the characteristics of features at different levels and reduce the loss of information.

In the diffusion process, the original cell embedding $x_{0}$ becomes a noised embedding $x_{T}$ by iteratively adding noise through $T$ steps. For the $i$-th step, the embedding $x_{i}$ is sampled from the following distribution:
\begin{equation}
q\left( x_{i} \middle| x_{i - 1} \right) = \mathcal{N}\left( x_{i} \middle| \sqrt{1 - \beta_{i}}x_{i - 1},\beta_{i}I \right), ~~\mathrm{where} \quad \beta_{i} \in (0,1)
\end{equation}
where $I$ stands for standard Gaussian noise. $\beta_{i}$ is a coefficient that varies with time step,
and $\beta_{min}$ and $\beta_{max}$ are two parameters that control the scale of $\beta_{i}$ in the diffusion process:
\begin{equation}
    \beta_{i} = \frac{\beta_{min}}{T} + \frac{i - 1}{T - 1}\left( {\frac{\beta_{max}}{T} - \frac{\beta_{min}}{T}} \right)
\end{equation}

The training goal is to learn the reverse diffusion process $p(x_{i-1}|x_i)$.
In each iteration, $x_{i-1}$ at step $i-1$ is predicted, given an embedding $x_i$ at step $i$. 
Such process also follows the Gaussian distribution. According to previous works \cite{ho2020denoising, dhariwal2021diffusion}, the mean and variance are parameterized as:
\begin{equation}\label{eq3}
    p_{\theta}(x_{i - 1} | x_{i}) = \mathcal{N}(x_{i - 1} | \mu_{\theta}(x_{i}, i),\mathrm{exp}({w \beta}_{i}){I})
\end{equation}
where $w$ in the variance is an adjustable weight that controls the randomness of the reverse process. The mean $\mu_{\theta}(x_{i}, i)$ can be written as:
\begin{equation}\label{eq4}
    \mu_{\theta}(x_{i}, i) = \frac{1}{\sqrt{\alpha_{t}}}\left(\mathbf{x}_{t}-\frac{\beta_{t}}{\sqrt{1-\bar{\alpha}_{t}}} \boldsymbol{\epsilon}_{\theta}\left(\mathbf{x}_{t}, t\right)\right)
\end{equation}
where $\alpha_{t} = 1-\beta_{t}$ and $\bar{\alpha}_{t} = \prod_{s=1}^t \alpha_{s}$. 
${\epsilon}_{\theta}\left( x_{i}, i \right)$ is the added noise predicted by the backbone network. In other words, the backbone network takes the cell's latent space embedding $x_{i}$ and the timestamp $i$ as inputs to predict the noise.



In the inference process, the diffusion model takes the Gaussian noise as the initial input and denoises it iteratively through T steps. Eventually, we can get the new cellular latent space embedding $x_0$ and put it into the decoder to get the final gene expression data.

\subsection{Conditional generation and the Gradient Interpolation strategy}
We use the classifier guidance method to perform conditional generation. 
This method does not interfere with the training of the denoising network model. Instead, the classifier is first trained separately by using condition labels like cell types and then provides gradients to guide cell generation. 
Here, we designed the cell classifier as a four-layer MLP. After generating a series of embeddings from cells with labels, the classifier takes both timestamp $i$ and cell embedding $x_{i}$ as inputs and predicts the cell labels $y$ {paired with $x_0$}. The cross-entropy loss is used for training. 
It is worth noting that only the embeddings between step $0$ and step $T/2$ of the diffusion process are used {for training the classifier}, considering that the signal in the rest part is too noisy to be predicted. 

As for inference, 
{given each step $i$ between the last part of the reverse process (between timestamp 0 and $T/2$)},
the classifier receives the intermediate state $x_i$ and outputs the predicted probability for every cell type. 
By computing the cross entropy loss between the predicted and desired condition given by the user, the gradient derived from the classifier can guide the denoising network model to generate a designated endpoint. The new embedding with the guidance is now sampled from:
\begin{equation} \label{eq5}
    p_{\theta}(x_{i - 1} | x_{i}, y) = \mathcal{N}(x_{i - 1} | \mu_{\theta}(x_{i}, I) + \beta_{i}\gamma\nabla_{x_{i}}{\log p_{\phi}}\left( {y \mid x_{i}} \right),\mathrm{exp}({w \beta}_{i}){I})
\end{equation}
where $p_{\phi}\left(y\mid x_i\right)$ stands for the classifier's result, and $\gamma$ is a weight that controls the effectiveness of the classifier to the reverse process. $\phi$ indicates the trainable parameters in the classifier. This guidance will affect every step of the reverse process and finally help the model's output reach a certain condition.

Since the classifier is trained aside from the diffusion model and is only used in the inference stage, we can train multiple classifiers \{$\phi_{1},\phi_{2},...$\} to control different conditions separately. The gradient that guides the diffusion process is the summation of all the classifiers' gradients with different weights \{$\gamma_{1},\gamma_{2},...$\}.



We proposed the Gradient Interpolation strategy to generate continuous cell condition guidance. A classifier receives two different conditions such as the initial and end state of cell differentiation, and generates two gradients at the same time.
These gradients are then integrated to guide the diffusion to an unseen intermediate state. Specifically speaking, the $\beta_{i}\gamma\nabla_{x_{i}}{\log p_{\phi}}\left( {y \mid x_{i}} \right)$ in Eq. \ref{eq5} is replaced by:
\begin{equation}
    \beta_{i}\gamma\nabla_{x_{i}}{\log p_{\phi}}\left( {y \mid x_{i}} \right) \rightarrow \beta_{i}(\gamma_{1}\nabla_{x_{i}}{\log p_{\phi}}\left( {y_{1} \mid x_{i}} \right)+\gamma_{2}\nabla_{x_{i}}{\log p_{\phi}}\left( {y_{2} \mid x_{i}} \right))
\end{equation}
where $\gamma_{1}$ and $\gamma_{2}$ represent two adjustable coefficients that control the distance between the generated cells and the two target cell states. 
By tuning these coefficients, scDiffusion can decide which cell state the generated cell is closer to, thus generating cells with continuous states. 
With this strategy, the initial state of the diffusion generation process is changed from pure Gaussian noise to the latent space embedding of cells of the initial condition, following a noise addition process:
\begin{equation} \label{eq7}
    x_{init}=\sqrt{\alpha_{t}}x_{0}+\sqrt{1-\alpha_{t}}\epsilon
\end{equation}
where $x_{init}$ is the initial state, and $t$ is a parameter that is smaller than the total diffusion step. $\alpha_{t}$ is the same thing as in Eq. \ref{eq4}. 
This modification preserves the general characteristics of the initial cells, allowing the model to generate a series of new cell states for each given initial state. These generated cells can constitute a continuous trajectory of cell states.


\subsection{Evaluation metrics}
\label{eval_metrics}
To compare the similarity between generated and real cells, we evaluated the generated data with various metrics. 
The statistical indicators consist of Spearman Correlation Coefficient (SCC), Maximum Mean Discrepancy (MMD) \cite{gretton2012kernel}, local inverse Simpson’s index (LISI) \cite{haghverdi2018batch}. and quantile-quantile plot (QQ-plot). 
We log-norm the gene expression data of generated and real cells and calculated SCC between them. 
The LISI score was calculated on the data-integrated KNN graph by using the Python package scib \cite{luecken2022benchmarking}. The QQ-plot was drawn using both real and generated expression data of a specific gene for a given cell type.


The non-satistical metrics include Uniform Manifold Approximation and Projection (UMAP) visualization \cite{mcinnes2018UMAP}, marker gene expression, CellTypist classification \cite{dominguez2022cross}, random forest evaluation, and KNN evaluation. The UMAP plot was used to visualize the generated and real expression data on a two-dimensional plane to provide a subjective judgment for the generated data. 
 
CellTypist \cite{dominguez2022cross} is used to judge whether the conditionally generated data can be classified into the right type. 
The random forest evaluation shares the same idea with scGAN, which uses a random forest model with 1000 trees and 5 maximum depths to distinguish cells from real and generated, and the more similar these two cells are, the closer the area under the receiver operating characteristic (ROC) curve (AUC) metric for random forests approaches 0.5. The KNN evaluation metric is the same as the random forests metrics except the classifier model is switched to the KNN model.

\section{Results}

We conducted four experiments to demonstrate the capability of scDiffusion. 
First, we investigated the data generation ability of scDiffusion and compared it with the deep learning based method scGAN and statistical learning based method scDesign3. 
We then assessed scDiffusion on a single-conditional generation task to generate specific cell types. 
Furthermore, we applied scDiffusion in a multi-conditional generation case with both cell types and organs as conditions and used it to generate new cells under an unseen condition which is out of the distribution of the training data. 
Lastly, we employed the Gradient Interpolation strategy to generate intermediate states in cell reprogramming.

We employed five single-cell transcriptomic datasets in these experiments.
The Human Lung Pulmonary fibrosis (PF) dataset \cite{habermann2020single} is a large scRNA-seq dataset that contains more than 110 thousand cells of human lungs with PF. 
Tabular Muris \cite{schaum2018single} is a large-scale single cell transcriptomic database of mice across 12 organs. 
The Tabular Sapiens \cite{the2022tabula} is the first-draft human cell atlas of nearly 500,000 cells from 24 organs of 15 normal human subjects, we selected six cell types from five of these organs to do the experiments.
The Waddington-OT dataset \cite{schiebinger2019optimal} is a cell reprogramming dataset of mouse embryonic fibroblasts (MEFs), containing cells with different timestamps during an 18-day reprogramming process. 
The PBMC68k dataset \cite{zheng2017massively} is a classical scRNA-seq dataset that contains 11 different cell types of human peripheral blood mononuclear cells (PBMCs). As the CD4+ T helper 2 cells had an extremely low number and could not be classified by Celltypist, we removed them for downstream analysis. For all five datasets, we filtered out cells with less than 10 expression counts and genes that expressed in less than 3 cells. In the Celltypist training, the data were split into training and testing sets with a ratio of 0.8 to 0.2, whereas the random forest and KNN models utilized splits of 0.75:0.25 and 0.7:0.3, respectively.


In all experiments, we set the diffusion step to 1000. The parameter $\gamma$ in Eq. \ref{eq5} was set to 2. The parameter $w$  in Eq. \ref{eq3} and Eq. \ref{eq5} was set to 0.5. The parameter $t$ in Eq. \ref{eq7} was set to 600.



\subsection{Realistic scRNA-seq data generation}
\begin{figure}[h]
  \begin{center}
    \includegraphics[width=1.0\textwidth]{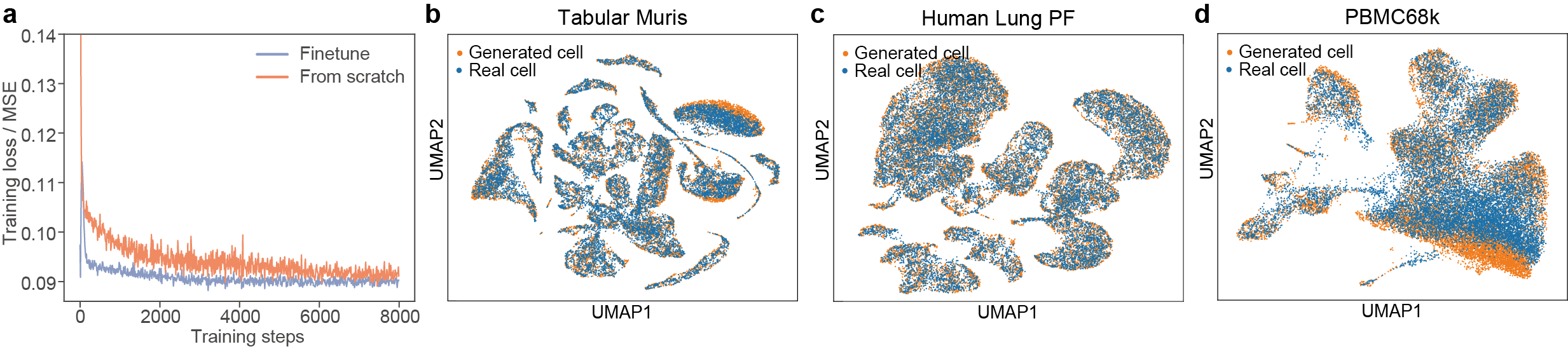}
  \end{center}
\caption{scDiffusion can generate realistic cell data. 
(a) The training loss curve of fine-tuning autoencoder from pre-trained SCimilarity weight and training autoencoder from scratch.
(b) UMAP of scDiffusion-generated Tabular Muris data and real Tabular Muris data. 
(c) UMAP of scDiffusion-generated Human Lung PF data and real Human Lung PF data. 
(d) UMAP of scDiffusion-generated PBMC68k data and real PBMC68k data. 
} 

\label{fig:result_noncondi}
\end{figure}

First, we tested the impact of the pre-trained model as autoencoder. We trained the SCimilarity model in two different ways, one was based on the pre-trained model parameters, and the other was from scratch. The result in Fig. \ref{fig:result_noncondi}a showed that the model based on the pre-trained weights could converge faster and be more accurate.
We then examined the ability of the diffusion model to generate without the guidance of the classifier, such ability is the foundation of the conditional generation. We applied scDiffusion on the Tabular Muris dataset, Human Lung PF dataset, and PBMC68k dataset to generate new cells (Fig. \ref{fig:result_noncondi}a, Fig. \ref{fig:result_noncondi} b and Fig. \ref{fig:result_noncondi} c). For comparison, we also generated cells with scGAN and scDesign3 using their default parameter settings.

We evaluated the performance of scDiffusion, scGAN, and scDesign with various metrics, and the results indicated that scDiffusion could generate realistic scRNA-seq data that was comparable with the state of the art method. The average SCC of scDiffusion in three datasets was 0.984, the mean MMD score was 0.018, the mean LISI score was 0.887 and the mean AUC of the random forest was 0.697, which are similar to the scGAN and scDesign3 as shown in the Fig. \ref{fig:static_metric}. These results provided a solid foundation for following conditional generation.

\subsection{Conditionally generating specific cell types}

We next trained a cell type classifier according to the annotations provided by the Tabular Muris dataset to guide the conditional generation of scDiffusion. For each cell type, we conditionally generated the same number of cells as the real data. 

As shown in Fig. \ref{fig:result_condi}a and Fig. \ref{condition_all}, the conditionally generated cells visually overlapped with the real cells on the UMAP plot. In order to compare the quality of conditionally generated cells, we trained the cscGAN model \cite{marouf2020realistic} with the same dataset and labels, conditionally generated each type of cell to make the comparison. Additionally, we added the cells generated by scDesign3 to the comparison since they have cell type labels, too.

We used Celltypist to classify these conditionally generated cells. 
As shown in \ref{table:classification_accuracies_muris}, the classification accuracies of the diffusion generated cells were close to those of the real cells in the test set, with the diffusion generated cells attaining an average accuracy of 0.93 across all cell types, compared to the 0.98 of the real cells. The cells generated by the scDesign3 also had high accuracies, achieving a mean accuracy of 0.99. In contrast, cells generated by the cscGAN could not be distinguished by the Celltypist, culminating in a mean accuracy of 0.04.

Considering that Celltypist was mainly used to distinguish between different cell types, we further used KNN model to distinguish between cells of each cell type and the corresponding real cells. The results in Fig. \ref{fig:result_condi}b showed that KNN models could not distinguish between diffusion-generated cells of a specified type and the real cells of that type, as the AUC scores in all kinds of cells are near 0.5, while it can easily distinguish the GAN-generated cells and scDesing3-generated cells, with basically all the AUC scores higher than 0.7.
We further examined the expressions of key genes in the tabular muris dataset. We selected five transcription factors (Klf13, Ybx1, Hnrnpk, Cnbp, Hmgb2) that have the highest mean Gini importance when making cell type classification in the original paper \cite{schaum2018single}, and drew the QQ-plots of these genes in different cell types between real and conditional generated data (Fig. \ref{qq_plot}a), which indicated that the expression of the key genes generated by the scDiffusion was close to that of real.

Similarly, we applied this analytical procedure to the PBMC68k dataset and observed congruent outcomes. (Fig. \ref{fig:result_condi}c, \ref{fig:result_condi}d and Fig. \ref{qq_plot}b). 
It's worth mentioning that rare cell types, such as the Thymus cell in the Tabular Muris dataset (2.5\% in the whole dataset) and the CD34+ cell in the PBMC68k dataset  (0.4\% in the whole dataset), can also be well generated. 

{
\begin{figure}[h]
  \begin{center}

    \includegraphics[width=0.95\textwidth]
    {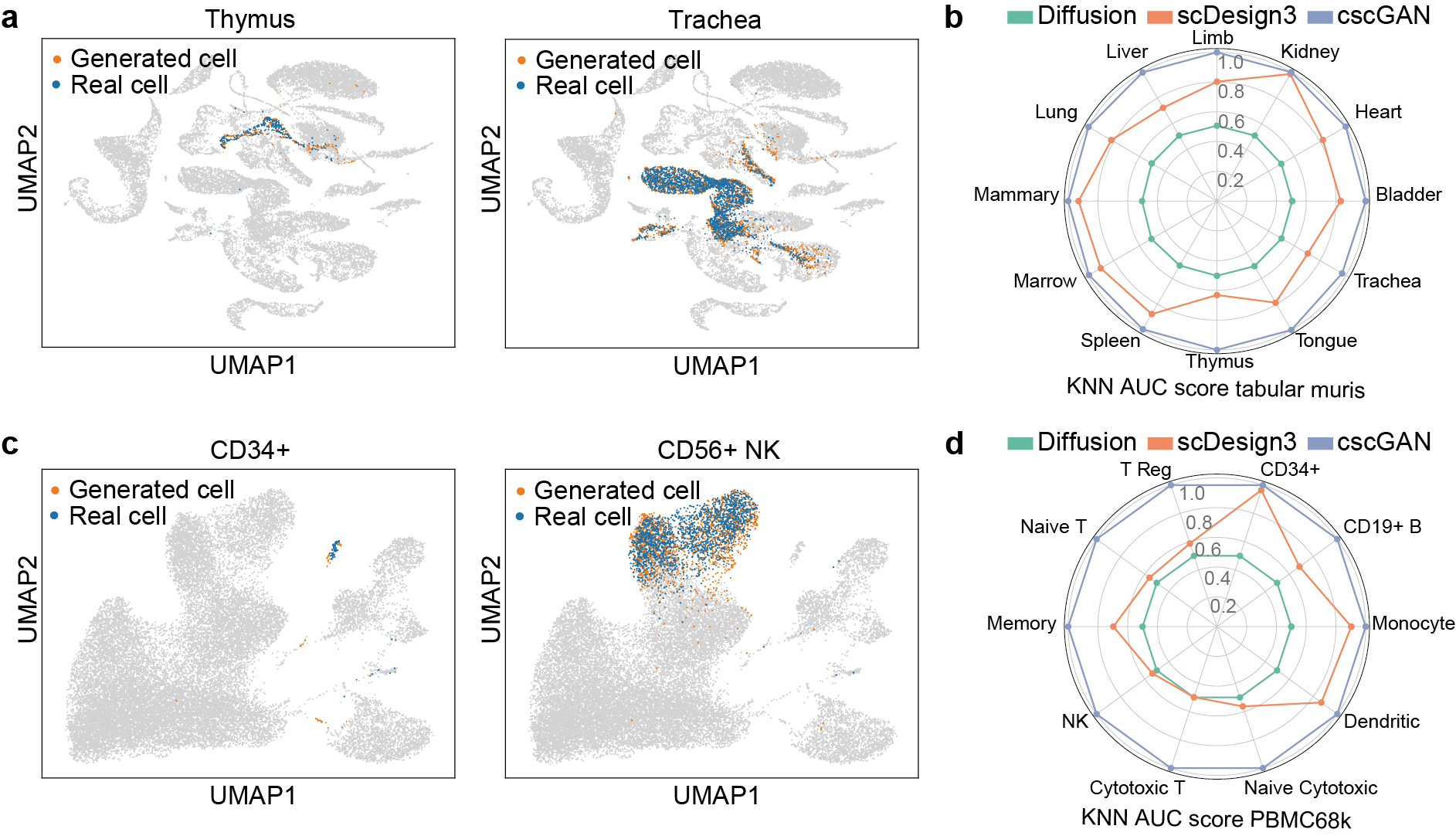}
  \end{center}
\caption{(a) UMAP of different cell types in the Tabular Muris dataset generated by conditional diffusion. The Thymus cell is a rare cell type. 
(b) The AUC score of KNN in different cell types in the Tabular Muris dataset. 
(c) UMAP of different cell types in the PPBMC68k dataset generated by conditional diffusion. The CD34+ cell is a rare cell type. 
(d) The AUC score of KNN in different cell types in the PBMC68k dataset.
}
\label{fig:result_condi}
\end{figure}
}

\subsection{Generating out-of-distribution cell data with multiple conditions}
We then tried to generate cells with multiple conditions based on the Tabular Muris dataset.
We trained two classifiers to separately control different conditions, one for organ type and the other for cell type.
We selected three cell groups, mammary gland T cell, spleen T cell, and spleen B cell, from the dataset for training. 
We would like to generate cells with a new combination of conditions (mammary B cell) which was not seen in the training data, or in other words, out of the distribution of the training data.



\begin{figure}[h]
  \begin{center}
    \includegraphics[width=1.0\textwidth]{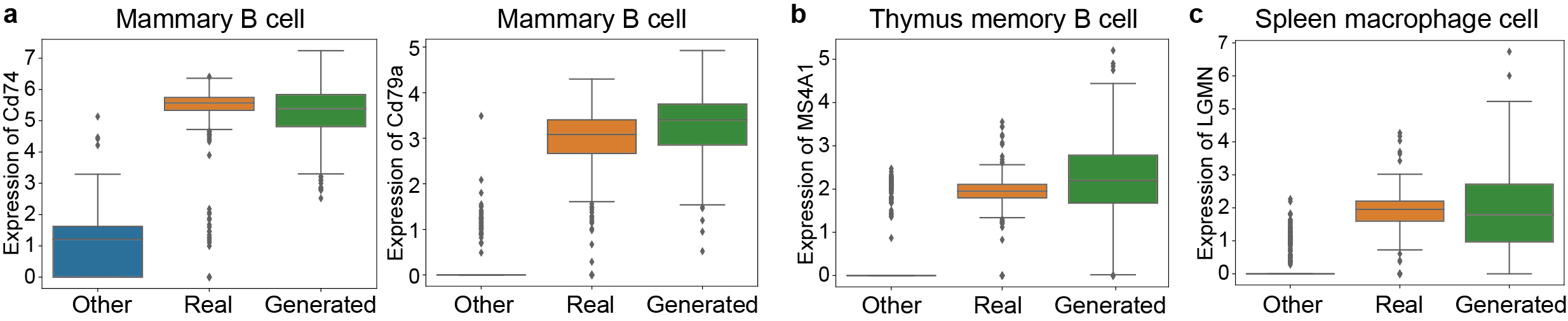}
  \end{center}
\caption{Marker genes' expression levels in real cells of this type, generated cells of this type, and real cells of other types. (a) Marker genes of mammary B cells. (b) Marker gene of thymus memory B cells. (c) Marker gene of spleen macrophage cells.}
\label{fig:ood}
\end{figure}

To test the generated multi-conditional data, we trained a Celltypist model with all kinds of cells in the mammary gland and used it to classify the real and generated mammary gland B cells. The result showed that 98\% of the generated cells and 92\% of the real B cells in the test set were categorized into the B cell, which showed that the scDiffusion can generate mammary B cells comparable to the real one. We then picked two marker genes of mammary B cells (CD74 \cite{bhatt2021startrac} and CD79A \cite{hilton2019single}) and drew the violin plot for them. The results in Fig. \ref{fig:ood}a showed that the generated marker genes have similar expression levels as the real ones.

We further selected cells from 5 organs and 6 cell types from the Tabular Sapiens dataset and did the same training process as above. The targets were changed to generate spleen thymus memory B cells and macrophage cells, which were removed from the training data. The results of the Celltypist showed that 96.75\% of the thymus memory B cells were categorized into memory B cells, and 96.63\% of the generated spleen macrophage cells were categorized into macrophage cells. As a comparison, the accuracy of real Thymus memory B cells and spleen macrophage cells were 96.91\% and 99.53\%. The violin plot of marker genes of thymus memory B cells (CD79A \cite{garman2020single}, SPIB, CD19 \cite{de2014purification} and MS4A1 \cite{nieto2021single}) and spleen macrophage cells (C1QB, C1QC \cite{missarova2021genebasis}, CD68 \cite{brown2005immunodetection} and LGMN \cite{zhao2020single}) also indicated similar levels of expressions (Fig. \ref{fig:ood}b, Fig. \ref{fig:ood}c and Fig. \ref{marker_rest}. These marker genes were selected using CellMarker2.0 \cite{hu2023cellmarker}.

All the results suggested that scDiffusion could effectively generate realistic out-of-distribution cells by learning the expression patterns of known cells.

\subsection{Generating intermediate cell states during cell reprogramming}

We used the Gradient Interpolation strategy to generate the intermediate cell states during cell reprogramming in the Waddington-OT dataset. 
We trained scDiffusion on the Waddington-OT dataset, which contains MEFs with the induction of reprogramming to induced pluripotent stem cells (iPSCs). The data were across 18 days since induction with a half-day interval, and a part of the cells were induced to redifferentiate at day 8.

\begin{figure}[hbt]
  \begin{center}
    \includegraphics[width=1.0\textwidth]{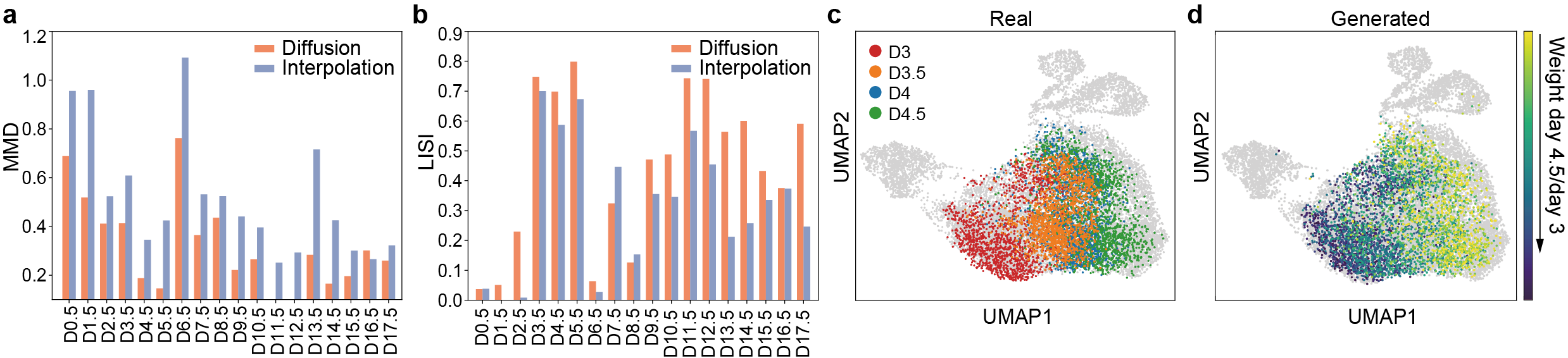}
  \end{center}
\caption{(a) The MMD score of different methods at different timestamps.
(b) The LISI score of different methods at different timestamps.
(c) UMAP of real cells.
(d) UMAP of cells generated by Gradient Interpolation. 
}
\label{fig:interpolate}
\end{figure}

We first trained scDiffusion with all integer days and generated cells in the middle of two integer days. We compared the results with direct interpolation. As some of the cells were induced to redifferentiate at day 8, we interpolated cells within each treatment group separately. 
The interpolation weights of both methods were set to 1:1. As shown in Fig. \ref{fig:interpolate}a and Fig. \ref{fig:interpolate}b, scDiffusion exhibited better performance than direct interpolation in MMD metrics and LISI metrics. The mean MMD and LISI of scDiffusion are 0.3217 and 0.4488, while the result of direct interpolation is 0.5206 and 0.3217, respectively. 
It is worth noting that scDiffusion was not trained with the information of different treatments, but its performance was still better than direct interpolation according to the treatment information, suggesting that the diffusion model can well capture the miscellaneous distribution of cells and well fit their intermediate states.

We then chose all samples from day 0 to day 8 with the exception of day 3.5 and day 4 to train scDiffusion, and trained the classifier with the same dataset using the timestamp as the label. 
We then sent two conditions, day 3 and day 4.5, to the classifier and used Gradient Interpolation to generate a series of cell states between day 3 and day 4.5 in the development trajectory (Fig. \ref{fig:interpolate}c). The initial noise was set to be the noised latent space embeddings of day-3 cells. 

We generated 10 states between day 3 and day 4.5 (Fig. \ref{fig:interpolate}d). We calculated the diffusion pseudotime \cite{haghverdi2016diffusion} of different states (Fig. \ref{fig:psudotime}) and found that state 6 and state 8 are the closest to real cells of day 3.5 and day 4, respectively. These cells were previously stripped out from the training data. 
We compared these two states with the direct interpolation of day 3 and day 4.5 whose weight was the same as Gradient Interpolation. The MMD scores of state 6 and state 8 are 0.047 and 0.0545, while the direct interpolation's scores are 0.1289 and 0.1167. The LISI scores of these two states with real day 3.5 and day 4 were 0.64 and 0.38, while the direct interpolations were 0.20 and 0.34. These results showed that scDiffusion can generate cells that are closer to the real intermediate state.


\section{Discussion}
  In this paper, we presented a deep generative neural network scDiffusion based on LDM and the foundation model. 
We used a pre-trained foundation model as an autoencoder and a skip-connected MLP backbone to make diffusion models suitable for gene expression data and generate realistic single cell data.
Using the classifier guidance method, scDiffusion could conditionally generate specific cell expression data based on user-defined conditions, including rare cell types. 
The flexibility of the classifier guidance also offers the potential to generate cells that are not seen in the training dataset. 
Furthermore, the Gradient Interpolation strategy enables the generation of a continuous cell trajectory between two known cell states to fill in the intermediate states. These abilities can be used to augment available scRNA-seq data and hold the potential for analyzing cell states that are not sequenced.

With its powerful generative ability, scDiffusion has the prospect of carrying on many other tasks. 
A very natural thing is multi-omics data generation. 
Theoretically, scDiffusion can generate any kind of single cell data. 
Besides, scDiffusion can also be used in the quality improvement of single cell data. 
For instance, by learning the overall expression paradigm in clean data, scDiffusion can perform denoising operations for contaminated data. 
In the future, we will try to replace the classifier with more powerful tools such as CLIP \cite{radford2021learning} in the stable diffusion \cite{rombach2022high}. In this way we may use more complex conditions to control the generating process and enable more complex tasks such as in silico cell perturbation, providing important help for drug selection and the control of cell state transition. The code of scDiffusion is available at https://github.com/EperLuo/scDiffusion.

\section{Acknowledgements} 
The work is supported in part by National Key R\&D Program of China (grant 2021YFF1200900), and National Natural Science Foundation of China (grants 62250005, 61721003, 62373210).

\bibliographystyle{myrecomb}

\bibliography{mybib}

\newpage





\centering{
\fontsize{18pt}{25pt}\selectfont\title{scDiffusion: conditional generation of high-quality single-cell data using diffusion model}
}

\author{  
Erpai Luo$^{1,^\#}$, Minsheng Hao$^{1,^\#}$, Lei Wei$^{1}$, Xuegong Zhang$^{1,2,^*}$
\\ \quad
\\
\fontsize{10pt}{\baselineskip}\selectfont $^{1}$MOE Key Lab of Bioinformatics and Bioinformatics Division of BNRIST, \\ 
\fontsize{10pt}{\baselineskip}\selectfont Department of Automation, Tsinghua University, Beijing 100084, China\\ 
\fontsize{10pt}{\baselineskip}\selectfont $^{2}$School of Life Sciences and School of Medicine, Tsinghua University, Beijing 100084, China\\	 
}

\date{}

\renewcommand{\thefootnote}{\fnsymbol{footnote}}  
\footnotetext{$^\#$ These authors contributed equally to this work.}  
\footnotetext{$^*$ Corresponding Author. Email: zhangxg@tsinghua.edu.cn} 

\maketitle

\centering
\section*{Supplementary Materials}  

\setcounter{figure}{0}
\renewcommand{\thefigure}{S\arabic{figure}}  

\begin{table}[ht]  
\centering  
\begin{tabular}{lcccc}  
\hline  
Tissue & Control Group & Diff Group & scDesign Group & GAN Group \\  
\hline  
Bladder & 0.987 & 0.987 & 1.000 & 0.000 \\  
Heart and Aorta & 0.961 & 0.665 & 1.000 & 0.000 \\  
Kidney & 0.985 & 0.915 & 1.000 & 0.000 \\  
Limb Muscle & 0.972 & 0.917 & 1.00 & 0.000 \\  
Liver & 0.989 & 0.992 & 1.000 & 0.000 \\  
Lung & 0.982 & 0.941 & 1.000 & 0.069 \\  
Mammary Gland & 0.966 & 0.899 & 1.000 & 0.247 \\  
Marrow & 0.987 & 0.953 & 0.998 & 0.016 \\  
Spleen & 0.991 & 0.996 & 0.997 & 0.066 \\  
Thymus & 0.956 & 0.925 & 1.000 & 0.090 \\  
Tongue & 1.000 & 0.996 & 1.000 & 0.000 \\  
Trachea & 0.998 & 0.983 & 1.000 & 0.001 \\  
\hline  
\end{tabular}  
\vspace{1.5em}
\caption{Comparison of classification accuracies for different cell types across four groups.}  
\label{table:classification_accuracies_muris} 
\end{table}

\begin{table}[ht]  
\centering  
\begin{tabular}{lcccc}  
\toprule  
Cell Type & Control Group & Diff Group & scDesign Group & GAN Group \\  
\midrule  
CD14+ Monocyte & 0.914 & 0.834 & 0.930 & 0.000 \\  
CD19+ B & 0.748 & 0.902 & 0.993 & 0.000 \\  
CD34+ & 0.936 & 0.810 & 1.000 & 0.000 \\  
CD4+/CD25 T Reg & 0.500 & 0.480 & 0.714 & 0.000 \\  
CD4+/CD45RA+/CD25- Naive T & 0.238 & 0.000 & 0.622 & 0.000 \\  
CD4+/CD45RO+ Memory & 0.264 & 0.188 & 0.651 & 0.000 \\  
CD56+ NK & 0.886 & 0.944 & 0.971 & 0.000 \\  
CD8+ Cytotoxic T & 0.682 & 0.772 & 0.775 & 0.003 \\  
CD8+/CD45RA+ Naive Cytotoxic & 0.719 & 0.965 & 0.707 & 0.998 \\  
Dendritic & 0.559 & 0.609 & 0.980 & 0.000 \\  
\bottomrule  
\end{tabular}  
\vspace{1em}
\caption{Comparison of classification accuracies for different cell types across four models in PBMC68k dataset.}  
\label{table:classification_accuracies_pbmc}  
\end{table}

\begin{figure}[h]
  \begin{center}
    \includegraphics[width=0.9\textwidth]{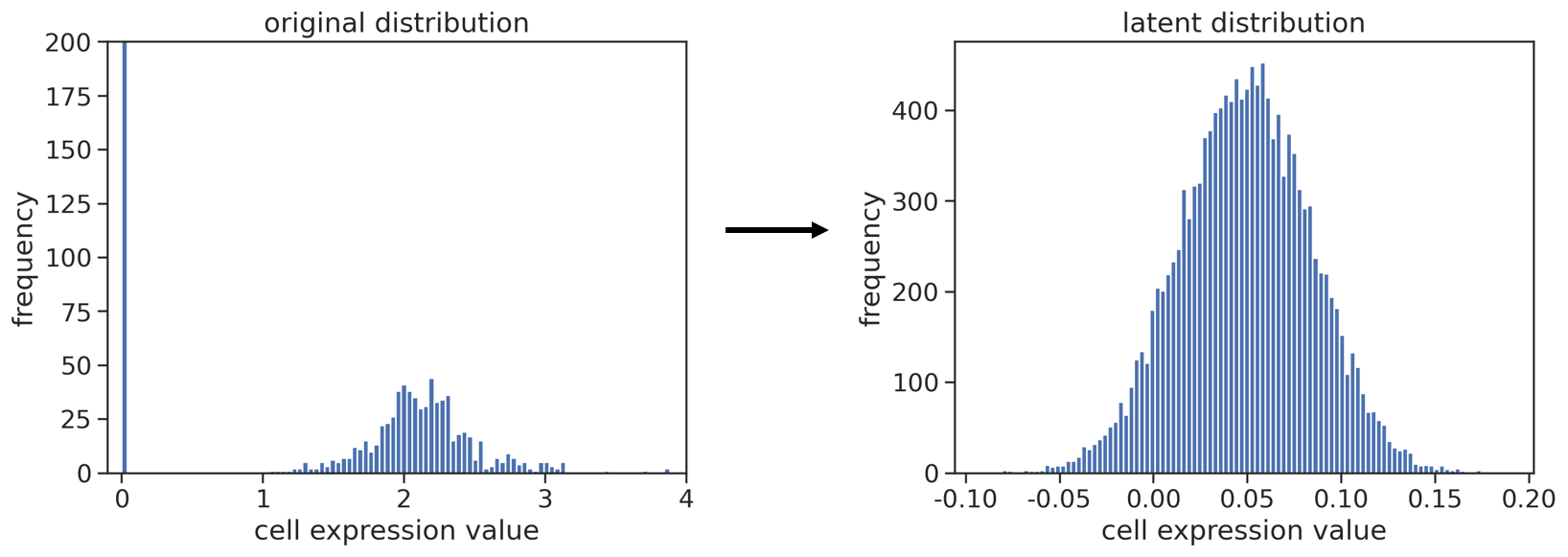}
  \end{center}
\caption{Distribution of original gene expression and latent embeddings derived by the autoencoder.}
\label{fig:dis_trans}
\end{figure}

\begin{figure}[h]
  \begin{center}
    \includegraphics[width=1.0\textwidth]{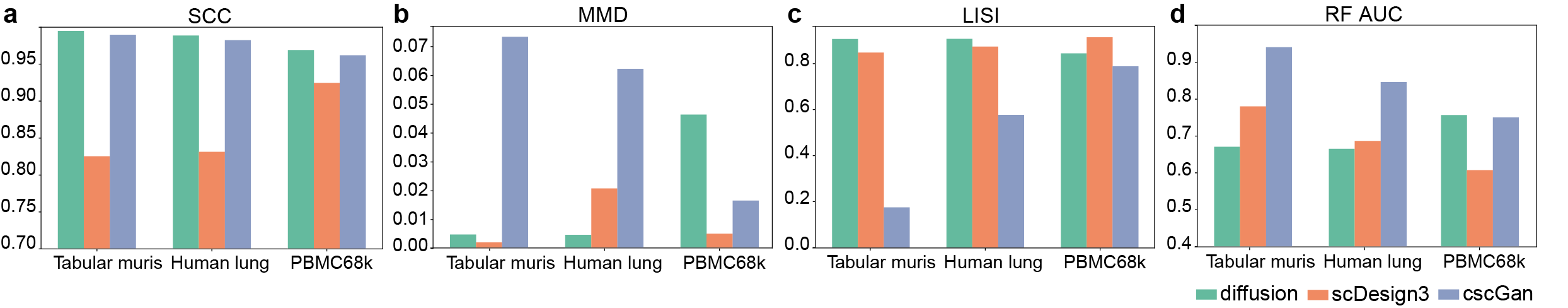}
  \end{center}
\caption{Different evaluation metrics to evaluate the realistic of generated cells in different methods.}
\label{fig:static_metric}
\end{figure}


\begin{figure}[h]
  \begin{center}
    \includegraphics[width=1.1\textwidth]{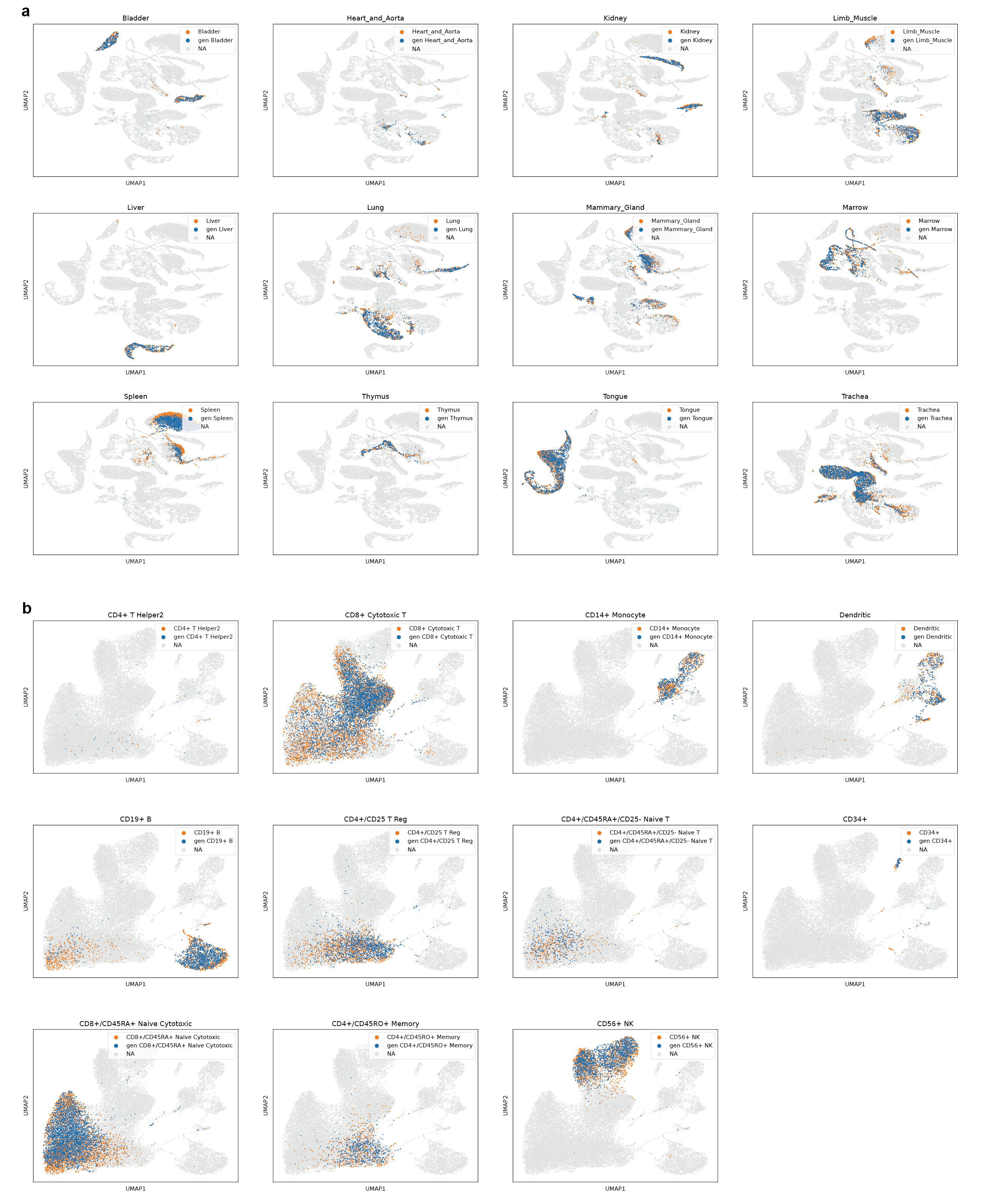}
  \end{center}
\caption{UMAP of conditionally generated cells. (a) The Tabular Muris dataset. (b) The PBMC68k dataset.}
\label{condition_all}
\end{figure}

\begin{figure}[h]
  \begin{center}
    \includegraphics[width=1.05\textwidth]{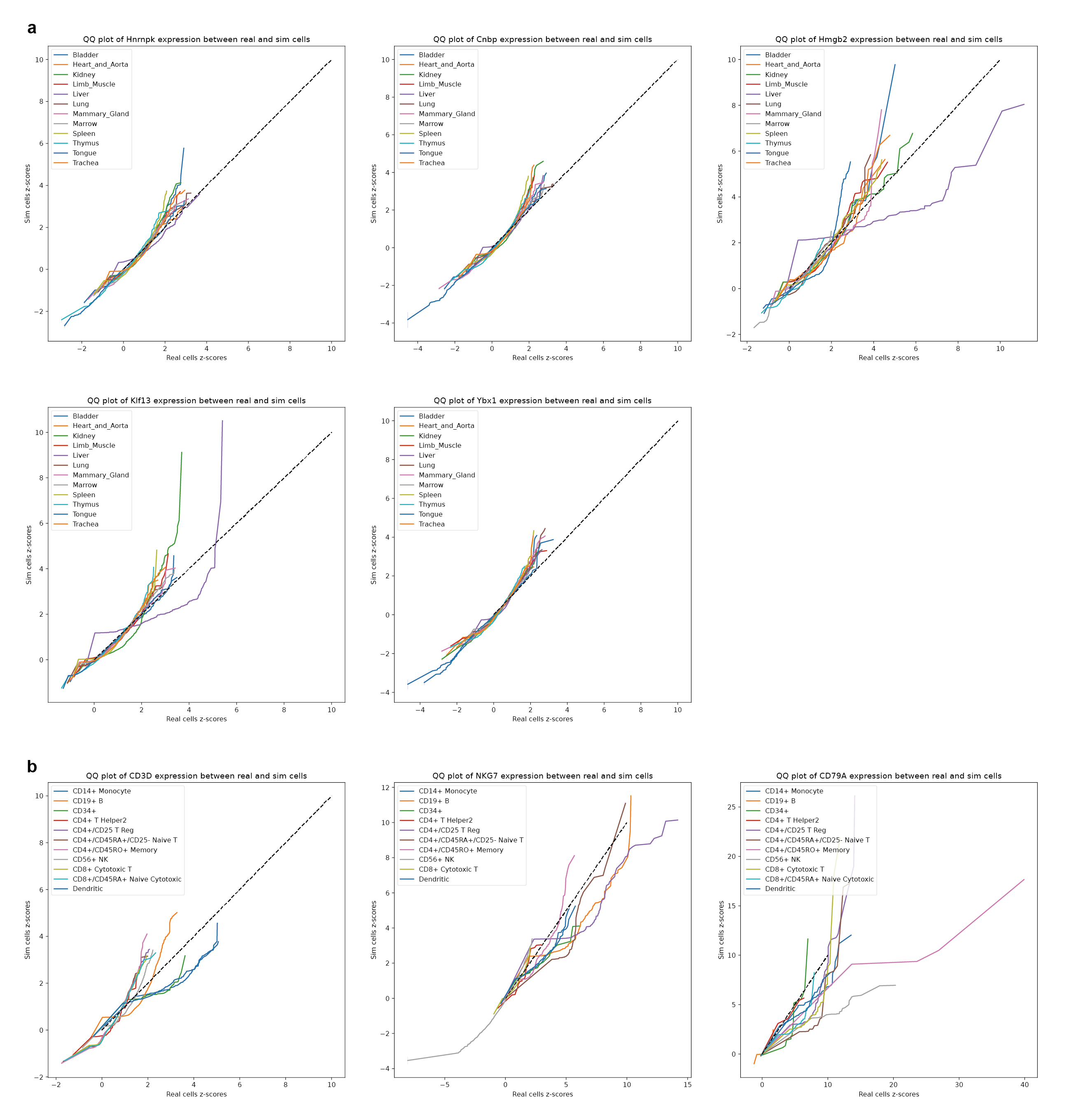}
  \end{center}
\caption{
QQ-plots of expression of feature genes in the real and generated data. (a) The Tabular Muris dataset. (b) The PBMC68k dataset.}
\label{qq_plot}
\end{figure}

\begin{figure}[h]
  \begin{center}
    \includegraphics[width=1\textwidth]{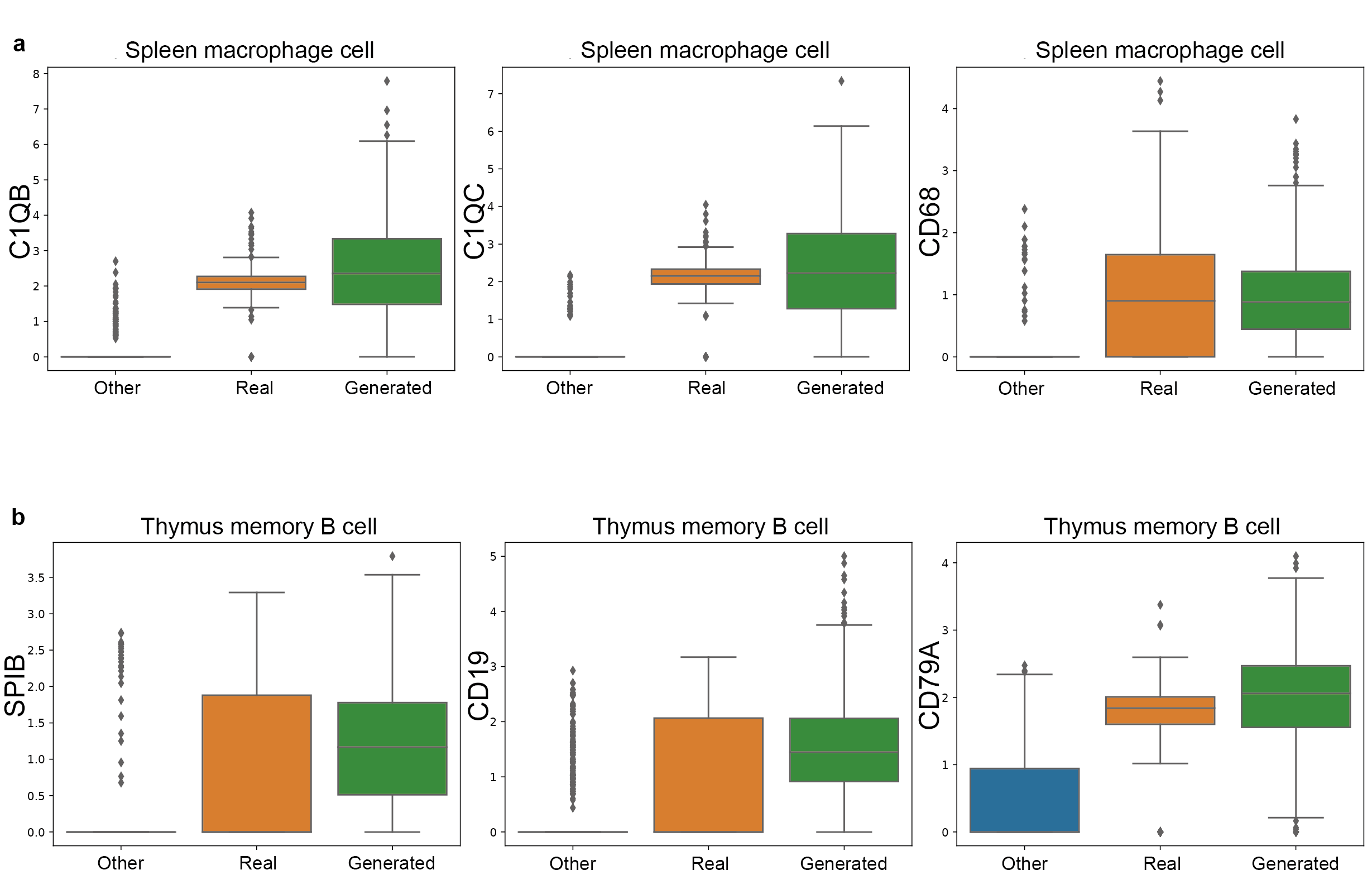}
  \end{center}
\caption{Significance of marker genes. (a) Maker genes of spleen macrophage cells. (b) Maker genes of Thymus memory B cells.}
\label{marker_rest}
\end{figure}

\begin{figure}[h]
  \begin{center}
    \includegraphics[width=0.7\textwidth]{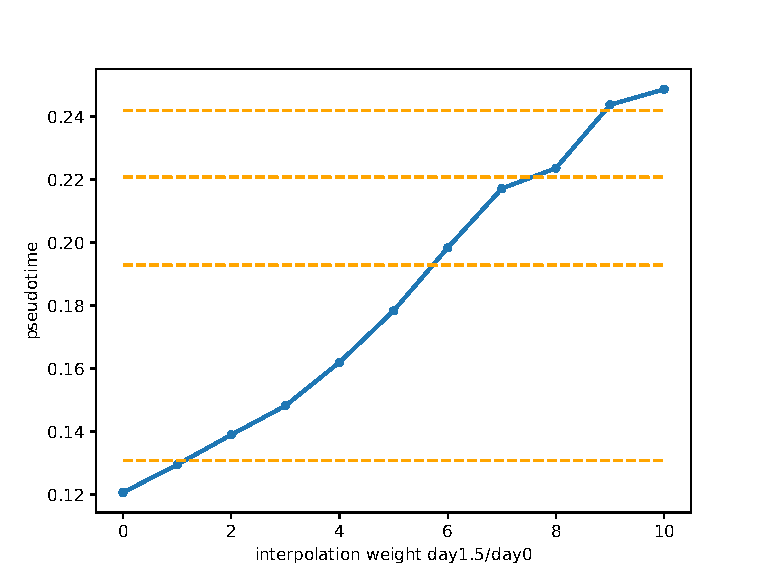}
  \end{center}
\caption{Pseudotime distance of generated states with different interpolation weights. Orange lines are the pseudotime of days 0, 0.5, 1, and 1.5 in the real data, from bottom to up.}
\label{fig:psudotime}
\end{figure}

\end{document}